\documentclass{article}
\usepackage[a4paper, total={6in, 8in}]{geometry}
\usepackage{graphicx}
\usepackage{booktabs}
\usepackage{caption}
\usepackage{tabularx}
\usepackage{float}
\usepackage{amsfonts, amsmath, amsthm, amssymb}
\usepackage{float}
\usepackage{xcolor}
\usepackage{titlesec}
\usepackage{caption}
\usepackage[utf8]{inputenc}
\usepackage{siunitx}
\usepackage{wrapfig}
\usepackage[
    natbib=true,
    date = year,
    style=authoryear,
    sorting=nyt,
    maxcitenames=1
]{biblatex}
\usepackage{geometry}
\title{Methods for Estimating the Exposure-Response Curve to Inform the New Safety Standards for Fine Particulate Matter}
\author{Michael Cork, Daniel Mork, Francesca Dominici}
\date{}

\begin{document}
  
\maketitle

\bigskip

\begin{abstract}
Exposure to fine particulate matter ($PM_{2.5}$) poses significant health risks and accurately determining the shape of the relationship between $PM_{2.5}$ and health outcomes has crucial policy ramifications. While various statistical methods exist to estimate this exposure-response curve (ERC), few studies have compared their performance under plausible data-generating scenarios. This study compares seven commonly used ERC estimators across 72 exposure-response and confounding scenarios via simulation. Additionally, we apply these methods to estimate the ERC between long-term $PM_{2.5}$ exposure and all-cause mortality using data from over 68 million Medicare beneficiaries in the United States. Our simulation indicates that regression methods not placed within a causal inference framework are unsuitable when anticipating heterogeneous exposure effects. Under the setting of a large sample size and unknown ERC functional form, we recommend utilizing causal inference methods that allow for nonlinear ERCs. In our data application, we observe a nonlinear relationship between annual average $PM_{2.5}$ and all-cause mortality in the Medicare population, with a sharp increase in relative mortality at low PM2.5 concentrations. Our findings suggest that stricter $PM_{2.5}$ limits could avert numerous premature deaths. To facilitate the utilization of our results, we provide publicly available, reproducible code on Github for every step of the analysis.

\bigskip
\bigskip

keywords: air pollution, all-cause mortality, causal inference, exposure-response curve, fine particulate matter, simulation study
\end{abstract}

\newpage

\section{Introduction}
In 2019, air pollution contributed to an estimated 6 million deaths globally, accounting for nearly 12\% of the total global mortality \parencite{healtheffectsinstituteStateGlobalAir2020, murrayGlobalBurden872020}. As the leading environmental risk factor for premature mortality, air pollution contributes to a higher number of estimated deaths each year than traffic collisions \parencite{healtheffectsinstituteStateGlobalAir2020, murrayGlobalBurden872020, manisalidisEnvironmentalHealthImpacts2020}. Long-term exposure to ambient fine particulate matter ($PM_{2.5}$) is the largest driver of air pollution’s burden of disease worldwide \parencite{healtheffectsinstituteStateGlobalAir2020}. In the United States, more than 40\% of the population resides in counties with unhealthy levels (above 35 $\si{\micro\gram/\meter^3}$) of particle pollution \parencite{ameircanlungassociationStateAir20222022}. While the detrimental health effects of $PM_{2.5}$ are widely acknowledged, describing the shape of the relationship between particulate matter pollution and adverse health outcomes, known as the exposure-response curve (ERC), remains uncertain. Accurately characterizing the ERC, particularly at lower exposure levels, carries significant policy implications.

In 2021, guided by epidemiological findings regarding the relationship between fine particulate matter and mortality risk, the World Health Organization (WHO) released new Global Air Quality Guidelines. These guidelines recommend reducing the annual average limit of $PM_{2.5}$ from 10 $\si{\micro\gram/\meter^3}$ to 5 $\si{\micro\gram/\meter^3}$ \parencite{worldhealthorganizationWHOGlobalAir2021}. These guidelines are widely used by decision makers and this change will influence the air quality standards implemented by various international governing bodies. In the United States (US), the Environmental Protection Agency (EPA) relies on the estimated shape of the ERC between air pollution and health outcomes to determine whether to lower the National Ambient Air Quality Standards (NAAQS). Epidemiological evidence in the US has found health risks persist at $PM_{2.5}$ levels below the current NAAQS set to 12 $\si{\micro\gram/\meter^3}$ for annual exposure, suggesting the need for a lower standard \parencite{diAssociationShorttermExposure2017b, weiCausalEffectsAir2020, shiNationalCohortStudy2021, ward-cavinessLongTermExposure2021, epaIntegratedScienceAssessment2019}. In January 2023, the EPA solicited public comment on a proposal to lower the NAAQS for annual $PM_{2.5}$ concentrations. The EPA estimates that adopting stricter standards could prevent up to 4,200 premature deaths annually \parencite{davenportBidenAdministrationMoves2023}. Recent evidence suggests that the impact of $PM_{2.5}$ may vary among marginalized subpopulations, highlighting the potential benefits of lower $PM_{2.5}$ levels for these communities \parencite{joseyAirPollutionMortality2023}. However, a significant challenge in lowering ambient air quality standards lies in establishing conclusive evidence of the causal link between particulate matter and health outcomes based on epidemiological data. Estimating the shape of the ERC and quantifying its uncertainty is complicated by the misaligned nature of the data, potential confounding factors, and effect heterogeneity.

To overcome these challenges, various methods have been proposed and implemented to characterize ERCs. In the field of air pollution epidemiology, ERCs are traditionally fit using a multivariate regression model with the health outcome as the dependent variable, air pollution exposure as an independent variable, and many potential confounders as additional independent variables \parencite{diAirPollutionMortality2017, liuAmbientParticulateAir2019}. Past research has often imposed an assumption of linearity between the natural log of the hazard ratio with respect to $PM_{2.5}$, restricting the ERC to represent an exponential increase in mortality per unit increase in pollutant concentration. Nonparametric or nonlinear functional regression methods provide greater flexibility in describing the shape of the ERC but may not adequately capture threshold effects \parencite{nasariClassNonlinearExposureresponse2016}. Other developed methods like the Extended Shape Constrained Health Impact Function (SCHIF) allow nonlinear association and can capture threshold behavior but impose more restrictive model specifications \parencite{burnettGlobalEstimatesMortality2018}. Additionally, some scientists argue against relying solely on regression-based ERC estimation in epidemiological studies, citing a lack of causal evidence, and advocate for the use of causal inference methods to inform air pollution policies \parencite{charteredcleanairscientificadvisorycommitteeSummaryMinutesEPA2018, goldmanDonAbandonEvidence2019}. Indeed, the EPA prioritizes studies employing causal inference methods for determining NAAQS limits \parencite{owensFrameworkAssessingCausality2017}.

Causal inference methods are often placed under the potential outcomes framework, which distinguishes between the design and analysis stages \parencite{imbensCausalInferenceStatistics2015}. In the design stage, researchers define the causal estimands and target population, and employ design-based methods like matching or weighting to construct a data set that emulates an experimental setup, where units with similar characteristics are compared across various exposure scenarios. After evaluating the design quality using metrics such as covariate balance, researchers proceed to the analysis stage to estimate causal effects. Causal inference methods, under certain assumptions, exhibit greater robustness to model misspecification compared to traditional regression methods \parencite{imbensCausalInferenceStatistics2015}. Nevertheless, many of the approaches for causal inference make the simplifying assumption of a binary exposure \parencite{robinsMarginalStructuralModels2000a, hernanMarginalStructuralModels2000, vanderlaanTargetedLearningCausal2011, rosembaumCentralRolePropensity1983, rubinMatchingUsingEstimated1996}. Initially, methods for estimating causal ERCs focused on weighting approaches utilizing the generalized propensity score (GPS) \parencite{robinsMarginalStructuralModels2000a, hiranoPropensityScoreContinuous2005, robinsEstimationRegressionCoefficients1994}. However, these methods are sensitive to GPS model misspecification and extreme weights. Doubly robust approaches mitigate this issue and provide asymptotically unbiased estimates of the ERC when either the outcome model or GPS model are misspecified \parencite{kennedyNonparametricMethodsDoubly2017, colangeloDoubleDebiasedMachine2022, schulzDoublyRobustEstimation2021}. Recently, weighting methods that directly optimize covariate balance have been extended to the continuous exposure setting \parencite{vegetabileNonparametricEstimationPopulation2021}. Another contemporary development focuses on extending the matching framework to the context of continuous exposures via a GPS caliper matching framework \parencite{wuMatchingGeneralizedPropensity2022}. Each proposed method has been tailored to address specific requirements in causal inference. However, limited research has compared the behavior of these methods under different assumed exposure-response relationships and confounding mechanisms.

In this paper, we aim to fill this research gap by conducting a comprehensive simulation study to compare various estimators of the exposure-response curve. We evaluate the performance of seven estimators, including both regression and causal inference methods, across a range of plausible ERC scenarios and confounding relationships. Furthermore, we apply these methods to estimate the ERC between long-term $PM_{2.5}$ exposure levels and all-cause mortality in a large observational administrative cohort comprising over 68 million Medicare beneficiaries in the continental United States from 2000 to 2016. Notably, this study represents the first application of multiple statistical approaches to a data set encompassing more than 500 million person-years of Medicare data. To ensure reproducibility and facilitate public access, we utilize methods available on CRAN and provide code for each step of the analysis on Github. The analysis code can be accessed at https://github.com/macork/ERC\_simulations, while the code detailing the data processing for our data application can be found at https://github.com/NSAPH/National-Causal-Analysis.

\section{Methods}
\subsection{Estimators and estimand}
We begin by providing a formal definition of our target estimand. Let $E_i \in \mathbb{R}^{+}$ represent a continuous, non-negative treatment or exposure for the $i^{th}$ unit in our target population. Adopting the potential outcomes framework and notation from Imbens and Hirano, \parencite{imbensRolePropensityScore2000, imbensNonparametricEstimationAverage2004} the random variable $Y_i(e)$ denotes the potential outcome for subject $i$ when exposed to a level $e \in \mathcal{E}$ of the exposure, where $\mathcal{E}$ denotes all levels of the exposure. We are interested in the random variable $Y_i(e)$, which represents the potential outcome under different levels of the exposure. However, note that $Y_i(e)$ cannot be directly estimated across various exposure values since $Y_i(e)$ is only observed for one exposure value for each unit, known as the fundamental problem in causal inference. Hence, we shift our target estimand to the population ERC $R(e)$, defined as the expected outcome at each exposure level $R(e) = E[Y_i(e)]$.

In this analysis, we compared the performance of seven different estimators in approximating $R(e)$ across various assumed exposure-response relationships and confounding mechanisms (Table 1). The first three estimators used to approximate the ERC are regression methods often used in practice. These regression methods included a linear model that adjusts for all variables in a linear manner (Linear), a generalized additive model that accounts for confounders linearly while allowing for a flexible nonlinear relationship between exposure and outcome through a spline basis representation (GAM), and a change point model that fits a segmented linear model with a single break point (Change point). The remaining four estimators are design-based methods commonly used in the field of causal inference. Specifically, we employed the same linear, GAM, and change point outcome models. However, instead of adjusting for confounding using linear regression terms, we utilized entropy balancing to generate continuous treatment weights. Entropy balance weights possess the desired characteristic of yielding nonzero weights while avoiding extreme values that may introduce bias \parencite{tubbickeEntropyBalancingContinuous2022}. We balanced the first moment of each covariate with the exposure, ensuring the covariates and exposure were uncorrelated in the weighted sample. To mitigate the presence of extreme weights, we ran the entropy balancing algorithm twice. First, we generated weights and truncated the upper 0.5\% of weights to the 99.5\% percentile. Then, we reran the algorithm using the truncated weights as base weights to obtain our final entropy weights \parencite{vegetabileNonparametricEstimationPopulation2021}. Once we obtained the weights, we estimated the ERC using the same regression techniques based on the weighted sample (Linear entropy, GAM entropy, and Change point entropy). The seventh and final estimator, CausalGPS, extends matching to the context of a continuous exposure via a GPS caliper matching framework \parencite{wuMatchingGeneralizedPropensity2022}. Further details regarding this estimator can be found in the CausalGPS package and the supplementary materials.

\begin{table}[H]
    \captionsetup{justification=raggedright,singlelinecheck=false, font = large}
    \label{table1}
    \begin{tabularx}{\textwidth}{ l X }
        \toprule
        Estimator name   & Description  \\
        \midrule
        Linear           & Gaussian linear model that linearly adjusts for $E$ and $\mathbf{C}$   \\
        GAM           & Generalized linear model that uses a spline with four degrees of freedom to nonparametrically adjust for our exposure and linearly adjusts for $E$ and $\mathbf{C}$ \\
        Change point     & Generalized linear model with a threshold and linear adjustment for $\mathbf{C}$   \\
        Linear entropy        & Linear model with entropy based weight adjustment   \\
        GAM entropy       & Spline model with entropy based weight adjustment  \\
        Change point entropy     & Change point model with entropy based weight adjustment   \\
        GPS matching (CausalGPS) & Causal inference method that uses GPS matching in a continuous setting \\
        \bottomrule
    \end{tabularx}
    \caption{Description of estimators used to characterize ERC}
    \end{table}

\subsection{Simulation setup}
We conducted a comprehensive set of simulations to evaluate the performance of our methods in estimating the effect of a continuous exposure on a continuous outcome under various data generating mechanisms. In particular, we examined how each estimator performed in different scenarios: (1) when the marginal relationship between exposure and outcome changed, (2) when the relationship between the confounders and exposure shifted, (3) when there was an interaction between the exposure and confounders in the outcome model, and (4) when the sample size varied.

\subsubsection{Simulation settings}
For each observation $i = 1, \ldots, n$ we generated a vector of six covariates $\mathbf{C}_i =(C_{1i}, C_{2i}, ..., C_{6i})$ with five continuous components and one categorical component:
$$ (C_{1i}, ..., C_{4i})' \sim \mathcal{MVN}(\mathbf{0}, I_4), C_{5i} \sim V \: \{-2, 2 \}, C_{6i} \sim U \: (-3, 3) $$
where $\mathcal{MVN}(0, I_4)$ denotes a multivariate normal distribution, $I_4$ is the identity matrix, $V \: \{-2, 2 \}$ denotes a discrete uniform distribution, and $U(-3, 3)$ denotes a continuous uniform distribution. To generate the exposure $E_i$, we outlined four specifications of the relationship between the confounders and the exposure, which relied on the cardinal function $\gamma(\mathbf{C}) = -0.8 + (0.1,0.1,-0.1,0.2,0.1,0.1) \mathbf{C}$. The coefficients of the cardinal function $\gamma(\mathbf{C})$ were similar to those used in previous analyses \parencite{kennedyNonparametricMethodsDoubly2017,wuMatchingGeneralizedPropensity2022}. The relationship between the covariates and the exposure (exposure model) varied in terms of error distribution and complexity, as described by the following equations.

\begin{align*}
     E_{\text{linear}} &= 9 * \gamma(\mathbf{C})+ 18 + N(0,10) \\
     E_{\text{heavy-tail}} &= 9 * \gamma(\mathbf{C})+ 18 + \sqrt{5} \times T(3) \\
     E_{\text{nonlinear}} &= 9 * \gamma(\mathbf{C}) + 2 * C_{3}^{2} + 15 + N(0,10) \\
     E_{\text{interaction}} &= 9 * \gamma(\mathbf{C}) + 2 * C_{3}^{2} +  2 * C_{1}C_{4} + 15 + N(0,10)
\end{align*}

In the first scenario, denoted as $E_{linear}$, a linear relationship between $E$ and $\mathbf{C}$ is assumed, with the addition of normal noise characterized by $N(0,10)$, representing a normal distribution with a variance of 10. In the $E_{heavy-tail}$ scenario, the relationship between $E$ and $\mathbf{C}$ remains linear, but the exposure values were heavy-tailed and include extreme values. Specifically, we use $T(3)$ to represent a Student's t-distribution with 3 degrees of freedom. The $E_{nonlinear}$ scenario introduced a non-linear relationship between $E$ and $\mathbf{C}$ while retaining normal noise. In the fourth scenario, $E_{interaction}$, an interaction term was added to the relationship between the exposure and the confounders. The scale (9) and location (18, 18, 15, 15) parameters were specified to ensure that more than 95\% of the exposure values exist between the range of [0, 20]. These simulation scenarios were designed to be comparable to the exposure range observed in our data application. As a final step, we enforced that all exposure values be nonnegative to reflect plausible air pollution exposure values and ensure the validity of all the data generating mechanisms. This was achieved by oversampling the number of exposure values, allowing us to remove any negative values and then randomly sample the remaining exposure values to match the desired sample size.

In all scenarios, we generated the outcome variable $Y_i$ from a normal distribution $Y|E,C \sim N(\mu(E, C), 10^2)$, where the mean function $\mu(E, C)$ is determined by the exposure ($E$) and the confounders ($C$), and the standard deviation $\sigma$ is set to 10. We considered three different specifications of the exposure-response curves: linear, sublinear, and threshold, denoted as $\mu_{\text{linear}}$, $\mu_{\text{sublinear}}$, and $\mu_{\text{threshold}}$, respectively. These three ERCs represent plausible relationships between air pollution and adverse health outcomes. For each specification of the ERC, we examined two outcome models. The first model, denoted as $\mu_{\text{ERC}}$, assumed a linear relationship between the confounders and the outcome, independent of the exposure. The second model, denoted as $\mu_{\text{ERC,int}}$, incorporated an interaction term between the confounders and the exposure. This interaction term accounts for the presence of heterogeneous treatment effects, where the impact of the exposure on the outcome is influenced by an individual's covariate values. The formulations of the mean functions are provided as follows:

\begingroup 
\addtolength\jot{2pt} 
\begin{align*}
\begin{split}
  \mu_{\text{linear}}(E, C) &= 20 - (2, 2, 3, -1, 2, 2)*\mathbf{C} + E  \\
  \mu_{\text{sublinear}}(E, C) &= 20 - (2, 2, 3, -1, 2, 2)*C + 3\log(E + 1) \\
  \mu_{\text{threshold}}(E, C) &= 20 - (2, 2, 3, -1, 2, 2)*C + 1.5E[E > 5]  \\
  \mu_{\text{linear, int}}(E, C) &= \mu_{\text{linear}} + E(-0.1C_1 + 0.1C_3^2 + 0.1C_4 + 0.1C_5)\\
  \mu_{\text{sublinear, int}}(E, C) &= \mu_{\text{sublinear}}
   + 3\log(E + 1)(-0.1C_1 + 0.1C_3^2 + 0.1C_4 + 0.1C_5)\\
   \mu_{\text{threshold, int}}(E, C) &= \mu_{\text{threshold}} + 1.5E[E > 5](-0.1C_1 + 0.1C_3^2 + 0.1C_4 + 0.1C_5)
\end{split} 
\end{align*}
\endgroup

In total, we generated data for 72 different scenarios by combining four exposure specifications, six specifications of the mean function for our outcome model, and three sample sizes (N = 200, 1000, 10000). For each scenario, we conducted 100 simulations, where we simulated the exposure and outcome models in each iteration.

\subsubsection{Evaluation of estimators}
Under every data generating scenario we estimated the ERC using all seven estimators.  We evaluated the goodness of fit of the estimators by comparing the absolute bias and root mean squared error (RMSE) of the estimated ERCs. We estimated these two quantities empirically by first evaluating the metric at equally spaced values across the range $\hat{\mathcal{E}^*}$ for all simulation replicates. Then we averaged  the metrics for each point across the exposure range $\hat{\mathcal{E}^*}$. $\hat{\mathcal{E}^*}$ is a restricted support of $\hat{\mathcal{E}}$ where we excluded some mass at the boundary to avoid boundary instability and only evaluate the ERC from zero to twenty \parencite{kennedyNonparametricMethodsDoubly2017}. The two quantities are formally defined as:

\begin{equation}
    \begin{aligned}
        |\operatorname{Bias}| &=M^{-1} \sum_{m=1}^{M}\left|S^{-1}\sum_{s=1}^{S} \hat{R}_{s}\left(e_{m}\right)-R\left(e_{m}\right)\right| \\
        \operatorname{RMSE} &=M ^{-1} \sum_{m=1}^{M} \left( S ^{-1} \sum_{s=1}^{S}\left(\hat{R}_{s}\left(e_{m}\right)-R\left(e_{m}\right)\right)^2 \right)^{1/2}
    \end{aligned}
\end{equation}

\noindent Here, $M = 100$, and $e_1, . . . , e_M$ are equally spaced points in the restricted support $\hat{\mathcal{E}^*}$. $\hat{R}_s$ represents the estimate of the ERC in the $s^{th}$ simulation. We report the bias and RMSE for each approach averaged across $S = 100$ simulations.

For the causal estimators, we evaluated the quality of the design setup by assessing covariate balance. Covariate balance measures the similarity of the distribution of observed pre-exposure covariates across all exposure levels to avoid confounder bias. Previous literature suggests that achieving an absolute correlation of 0.1 or lower indicates empirical covariate balance \parencite{zhuBoostingAlgorithmEstimating2015}. In the case of entropy weighting, we calculated the absolute correlation by considering the weighted correlation between each covariate and the exposure, using the weights generated during the design phase. The CausalGPS method follows a similar procedure to compute the measure of absolute correlation \parencite{wuMatchingGeneralizedPropensity2022}.

\subsection{Data application}
We applied the proposed methods to estimate the effect of long-term PM$_{2.5}$ exposure on all-cause mortality using a previously identified data set \parencite{joseyAirPollutionMortality2023}. We utilized the cohort of Medicare enrollees across the contiguous US from 2000 to 2016, which includes demographic information such as age, sex, race/ethnicity, date of death, and residential ZIP code. The study population consisted of over 68 million individuals residing in 31,414 ZIP codes, for which we compiled the number of deaths among Medicare enrollees for each ZIP code and year. Annual estimates of $PM_{2.5}$ exposure were obtained from a validated ensemble prediction model developed in previous research \parencite{diEnsemblebasedModelPM22019}. To obtain the annual average PM$_{2.5}$ at each ZIP code, we aggregated the daily US PM$_{2.5}$ exposure estimates at a 1km x 1km grid cell resolution using area-weighted averages \parencite{diAirPollutionMortality2017}. We assigned the annual average PM$_{2.5}$ value to individuals residing in each ZIP code for each calendar year. The predicted annual average PM$_{2.5}$ ranged from 0.01 to 30.92 $\mu g/m^3$, with the 5th and 95th percentiles being 4.26 and 15.04, respectively. To avoid extrapolation at the boundaries of the exposure range, we trimmed the highest 5\% and lowest 5\% of PM$_{2.5}$ exposures in the ERC. This trimming practice aligns with previous literature \parencite{liuAmbientParticulateAir2019, diAirPollutionMortality2017} and assists in meeting the causal assumptions necessary for identifying our estimand of interest \parencite{petersenDiagnosingRespondingViolations2012}.

We employed each of the methods described in our simulation to estimate the ERC between $PM_{2.5}$ and all-cause mortality. The outcome variable used was the log-transformed mortality rate, while the covariates at the ZIP code level, along with the year and strata variables, were included in the models. To address zero rates, we replaced them with half the minimum observed mortality rate across all ZIP codes and years. Following the proposed methodology, we fit each model and exponentiated the mean estimates to obtain the estimated all-cause mortality rate as a function of PM$_{2.5}$ concentration. The entire ERC was reconstructed using estimates at 100 equidistant levels of exposure, ranging from the minimum to the maximum observed PM$_{2.5}$ concentration in our sample.

To address potential confounding factors, the data set incorporated 10 ZIP code- and county-level variables. These variables encompassed ZIP code-level socioeconomic status (SES) indicators obtained from the 2000 and 2010 Census and the 2005-2012 American Community Surveys (ACS), as well as county-level information from the Centers for Disease Control and Prevention's Behavioral Risk Factor Surveillance System (BRFSS) \parencite{wuEvaluatingImpactLongterm2020, joseyAirPollutionMortality2023}. Specifically, the potential confounders consisted of two county-level variables: average body mass index and smoking rate; and eight ZIP code–level census variables: proportion of Hispanic residents, proportion of Black residents, median household income, median home value, proportion of residents in poverty, proportion of residents with a high school diploma, population density, and proportion of residents that own their house. Additionally, we include four ZIP code–level meteorological variables: the summer (June to September) and winter (December to February) averages of maximum daily temperatures and relative humidity. The data set also included two indicator variables indicating (i) the four census geographic regions of the United States (Northeast, South, Midwest, and West) and (ii) calendar years (2000–2016) aiming to respectively adjust for any residual or unmeasured spatial and temporal confounding.

We aggregated the data set in this study at the ZIP code-year level and then stratified based on age, sex, Medicaid eligibility (as a proxy for individual-level socioeconomic status), and follow-up year. Since each ZIP code-year contained varying amounts of person-time contributing to each stratum, we employed a weighted regression with person-time as weights to accurately estimate the population ERC. However, we adapted the general approach for our causal inference estimators due to the misalignment between the exposure and outcome measurements. The exposure and confounder data were measured at the ZIP code-year level, whereas the mortality data incorporate demographic and structural strata within ZIP code-years. To address this discrepancy, we modified our approach, drawing on the work of Josey et al. \parencite{joseyAirPollutionMortality2023, balzerNewApproachHierarchical2019}. First, to account for confounding, we generated weights (entropy balancing weights for entropy weighting estimators and ``matching weights" for CausalGPS) that balanced the distribution of covariates across different exposure levels at the ZIP code-year level. This balancing mitigated confounding effects by breaking the association between exposure and covariates in the weighted data. Second, we specified an outcome model where we modeled mortality rates specific to ZIP code and year as a function of both individual-level and ZIP code-level covariates. Finally, we fit the outcome model by multiplying the balancing weights from the design stage of our analysis with the observed person-time within each ZIP code-year stratum. The validity of this weighting procedure has been outlined in the literature \parencite{dongUsingPropensityScore2020, zanuttoComparisonPropensityScore2006}. Further details regarding each step can be found in the supplementary materials.

We implemented the M-out-of-N bootstrap procedure, where M = N/log(N), to construct the point-wise Wald 95\% confidence band for the ERC \parencite{politisLargeSampleConfidence1994, bickelResamplingFewerObservations2012}. In this procedure, we utilized a block bootstrap with ZIP codes serving as the block units. This approach allowed us to consider the correlation between observations across different years but within the same ZIP code \parencite{wuMatchingGeneralizedPropensity2022}. For each bootstrap replicate, we recalculated the GPS and entropy weights and refit the outcome model. This ensured that the bootstrap procedure accounted for the variability associated with both the design and analysis stages of the causal inference estimators.

\newpage

\section{Results}
\subsection{Simulation results}
\subsubsection{Covariate balance}
Before assessing the results of our simulation, we first evaluated the performance of our causal inference estimators in achieving covariate balance. Covariate balance is determined by the relationship between the covariates and the exposure, and it should be similar across all outcome models in our simulation. Good covariate balance is demonstrated when the average absolute correlation between the covariates and the exposure is below 0.1 in the reweighted data set. In Supplementary Figure S1, we present the number of simulations that achieved covariate balance using either entropy balancing weights or the CausalGPS package. Entropy balancing weights consistently achieved covariate balance, regardless of the sample size, as they are designed to limit the absolute correlation. They only failed to achieve covariate balance in the heavy-tailed exposure setting under large sample sizes, which can be attributed to our truncation procedure. The GPS matching framework used in the CausalGPS package tended to achieve covariate balance in larger sample sizes (N = 10,000) or in the linear and heavy-tailed exposure models. On the other hand, it often failed to achieve balance in the nonlinear or interaction exposure scenarios at lower sample sizes. For instance, with a sample size of 200, only 21\% of simulations across all exposure models had a mean absolute correlation below 0.1 for the CausalGPS estimator, compared to 68\% and 93\% with sample sizes of 1,000 and 10,000, respectively. Additionally, in Supplementary Figure S2, we report the average and upper and lower bounds for the correlation achieved for each individual covariate using both entropy weights and the CausalGPS package, based on a sample size of 1,000. It is important to note that a mean absolute correlation below 0.1 does not guarantee that the absolute correlation falls below 0.1 for each individual covariate.

\subsubsection{Linear outcome model}
We begin by presenting our results for the linear ERC scenario ($\mu_{\text{linear}}, \mu_{\text{linear,int}}$). Figure 1 compares the absolute bias and RMSE of the seven estimators for a sample size of 1,000. Supplementary Figure S3 and S4 provide the absolute bias and RMSE across all sample sizes. In the absence of an interaction term ($\mu_{linear}$), all estimators exhibited relatively low bias. The linear model demonstrated the least bias and RMSE across different confounder settings. On the other hand, the CausalGPS package showed the highest bias and RMSE, which decreased with increasing sample size (Supplementary Figure S3).

In contrast, under an interaction outcome setting ($\mu_{linear, int}$), we observed that causal inference methods generally outperformed regression methods in terms of absolute bias. The entropy weights effectively debiased the results in the linear and heavy-tailed exposure settings. However, the use of entropy-based weights resulted in an increase in absolute bias when the exposure was defined by a nonlinear function of the covariates ($E_{\text{nonlinear}}, E_{\text{interaction}}$). In the nonlinear and interaction exposure settings, the CausalGPS package exhibited lower absolute bias compared to entropy-weighted or non-causal methods. Regarding RMSE, the causal inference methods generally yielded similar results to regression methods, except for the CausalGPS package, which showed higher RMSE. However, the RMSE of the CausalGPS package reached a level comparable to other estimators at larger sample sizes (Supplementary Figure S4). Supplementary Figure S5 displays the ERCs under the $\mu_{\text{linear,int}}$ model setting. These plots reveal that while the CausalGPS package exhibited minimal bias in its mean ERC curve for the nonlinear and interaction exposure setting, it displayed a higher level of variability. Additionally, the GAM and change point models underestimated the response at higher values of the exposure.

\begin{figure}[H]
    \centering
    \includegraphics[width=\textwidth]{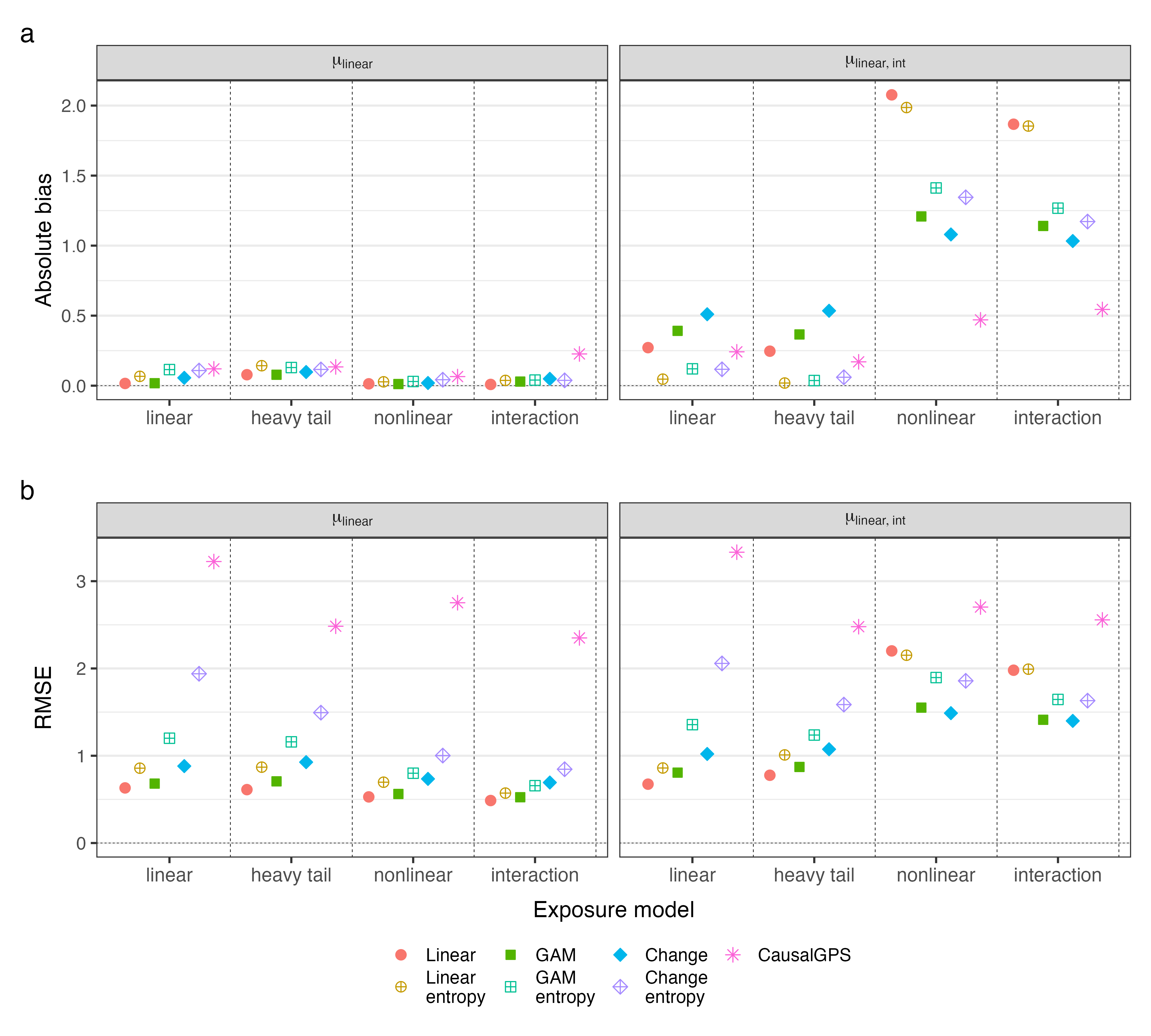}
    \caption{\textbf{Absolute bias and RMSE of ERC estimators in the linear outcome model setting}. (a) Absolute bias and (b) RMSE from simulations under different specifications of our estimator with sample size of 1000. We plot the mean absolute bias from 100 simulations. Plots are faceted on the x-axis by outcome model type (linear or interaction)}
\end{figure}

\subsubsection{Sublinear outcome model}
Next, we present results under the sublinear outcome model settings ($\mu_{\text{sublinear}}, \mu_{\text{sublinear,int}}$). In Figure 2 we compare the absolute bias and RMSE for the seven estimators under a sample size of 1000. In the sublinear outcome model setting without interaction ($\mu_{sublinear}$), the GAM, change point, and CausalGPS estimators demonstrated the least absolute bias. The GAM models consistently showed the lowest RMSE in this setting, while CausalGPS exhibited a large RMSE.

When considering the presence of heterogeneous treatment effects ($\mu_{sublinear, int}$), the GAM, change point, and CausalGPS estimators again resulted in the least absolute bias under the linear and heavy tail exposure settings. In the nonlinear and interaction exposure settings, entropy weighting did not de-bias the ERC, and only the CausalGPS package exhibited low bias. The CausalGPS estimator tended to de-bias more effectively at larger sample sizes (Supplementary Figure S6). Regarding RMSE, we observed that the GAM model consistently had the lowest RMSE in the $\mu_{\text{sublinear,int}}$ setting as well. This result is surprising given the known bias in the model. We observed that the CausalGPS package had a larger RMSE, which decreased with sample size (Supplementary Figure S7). Supplementary Figure S8 demonstrates that only CausalGPS captured the ERC in the nonlinear and interaction exposure setting when assessing the mean ERC at low or high exposure levels.

\begin{figure}[H]
    \centering
    \includegraphics[width=\textwidth]{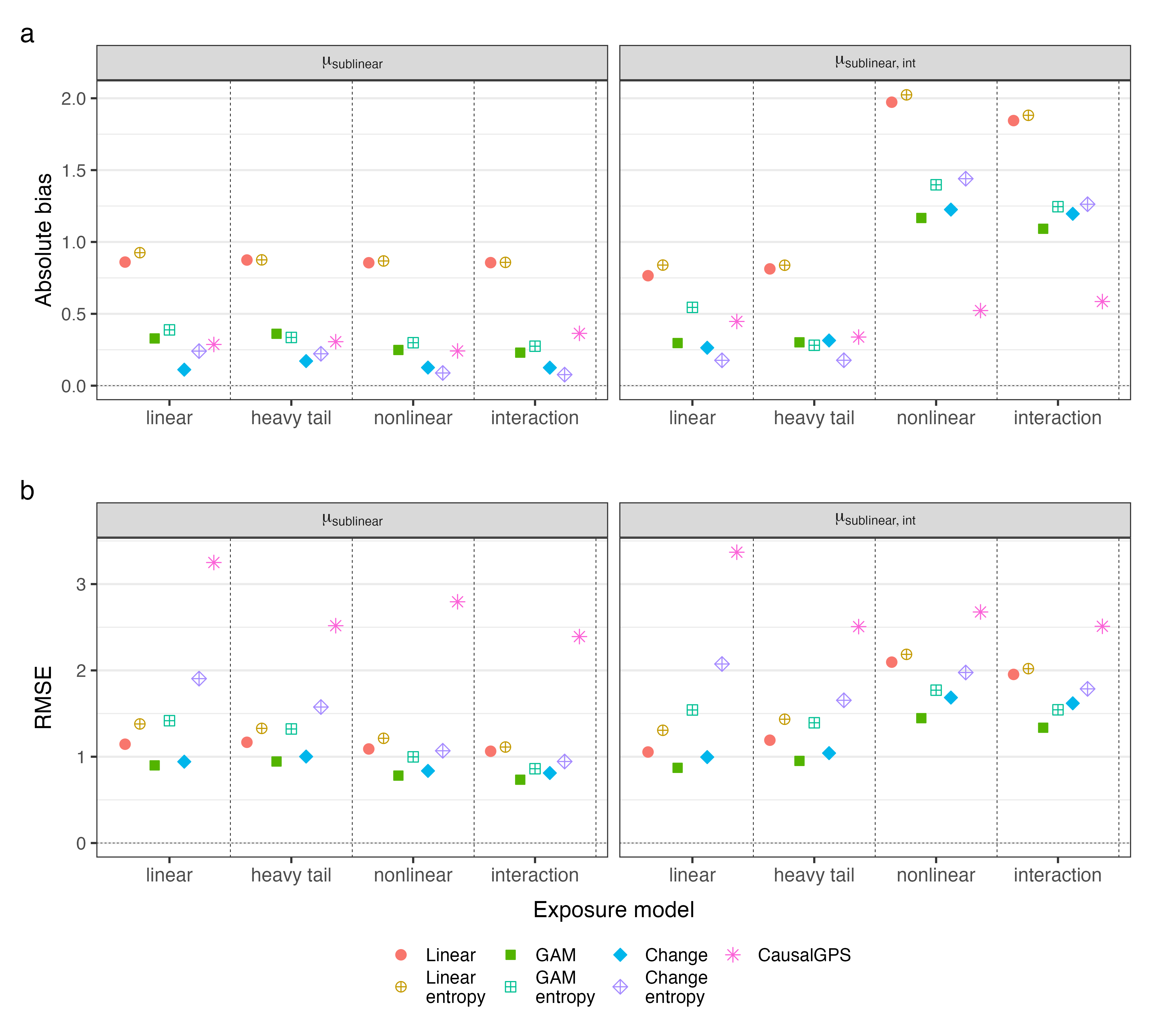}
    \caption{\textbf{Absolute bias and RMSE of ERC estimators in the sublinear outcome model setting}. (a) Absolute bias and (b) RMSE from simulations under different specifications of our estimator with sample size of 1000. We plot the mean absolute bias from 100 simulations. Plots are faceted on the x-axis by outcome model type (linear or interaction)}
    \label{fig1}
\end{figure}

\subsubsection{Threshold outcome model}
We computed the same metrics in the threshold outcome model setting (Figure 3). In the absence of interaction ($\mu_{\text{threshold}}$), we observed that the change point models, with and without entropy weighting, exhibited the least absolute bias. When assessing RMSE, both the change point models and GAM models performed well. At larger sample sizes, the entropy-weighted change point models showed comparable RMSE (Supplementary Figure S10).

Under an interaction mean outcome model ($\mu_{\text{threshold,int}}$), the Change point entropy model demonstrated the lowest absolute bias in the linear or heavy-tailed exposure setting. The CausalGPS package exhibited low absolute bias across all exposure settings. When comparing RMSE, the entropy weighting scheme generally led to increased RMSE compared to the unweighted estimators. However, this disparity disappeared at larger sample sizes (Supplementary Figure S10). The GAM and change point models tended to have the lowest RMSE across the four exposure scenarios. When assessing the ERCs in Supplementary Figure S11, few estimators captured the threshold behavior. CausalGPS captured no effect below 5 on average in the $E_{nonlinear}$ and $E_{interaction}$ settings and did not attenuate effects at higher exposure levels. The Change point entropy model captured the threshold only in the $E_{linear}$ and $E_{heavytail}$ settings.

\begin{figure}[H]
    \centering
    \includegraphics[width=\textwidth]{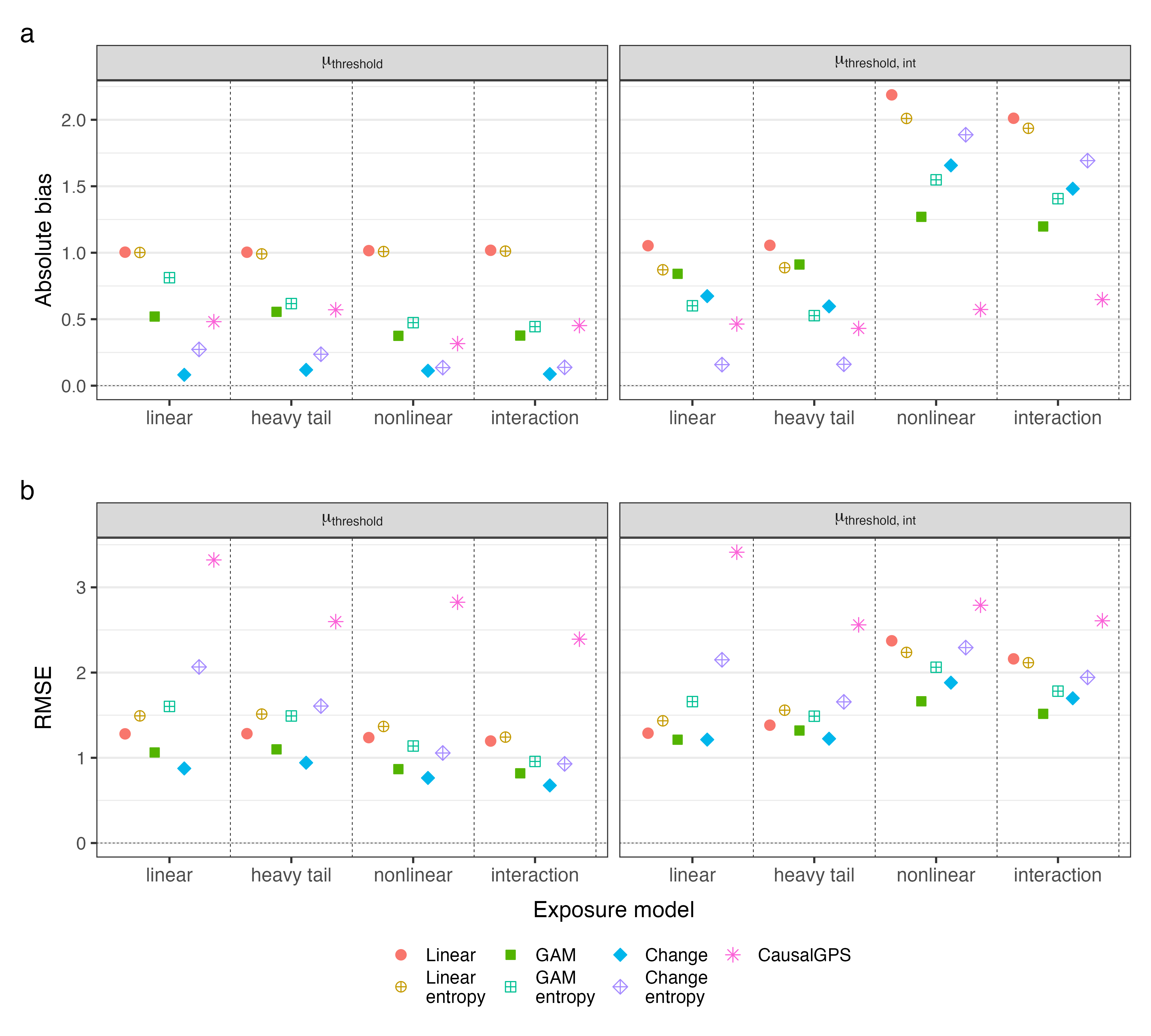}
    \caption{\textbf{Absolute bias and RMSE of ERC estimators in the threshold outcome model setting}. (a) Absolute bias and (b) RMSE from simulations under different specifications of our estimator with sample size of 1000. We plot the mean absolute bias from 100 simulations. Plots are faceted on the x-axis by outcome model type (linear or interaction)}
\end{figure}

\subsection{Data application}
We present the covariate balance plots for the causal inference estimators using the Medicare data in Figure 4. Note that while the absolute correlation between each covariate and the exposure was not below 0.1 for the CausalGPS estimator, the mean absolute correlation did fall below 0.1.

Figure 5 displays the estimated ERC for mortality rate as a function of annual average $PM_{2.5}$. The change point models, which failed to converge for this large data set, were omitted from the plot. Figure 5 demonstrates that nonlinear causal inference methods (GAM entropy, CausalGPS) exhibited a more pronounced increase in mortality at lower levels of annual average $PM_{2.5}$ before attenuating at higher concentrations, similar to our sublinear outcome model in the simulations.

Figure 6 presents the estimated relative mortality rates for the Medicare population. We compared the estimated mortality rate for each estimator to the estimated mortality rate at the current EPA limit of 12 $\si{\micro\gram / \meter^3}$. All estimators projected a decrease in relative mortality for the Medicare population at concentrations lower than 12 $\si{\micro\gram / \meter^3}$. The CausalGPS estimator demonstrated a more gradual decrease in relative mortality from 9-12 $\si{\micro\gram / \meter^3}$ compared to other estimators, followed by a sharp decrease at concentrations lower than 9 $\si{\micro\gram / \meter^3}$. The GAM, GAM entropy, Linear, and Linear entropy estimators illustrated similar decreases with overlapping confidence intervals.

\begin{figure}[H]
    \centering
    \includegraphics[width=\textwidth]{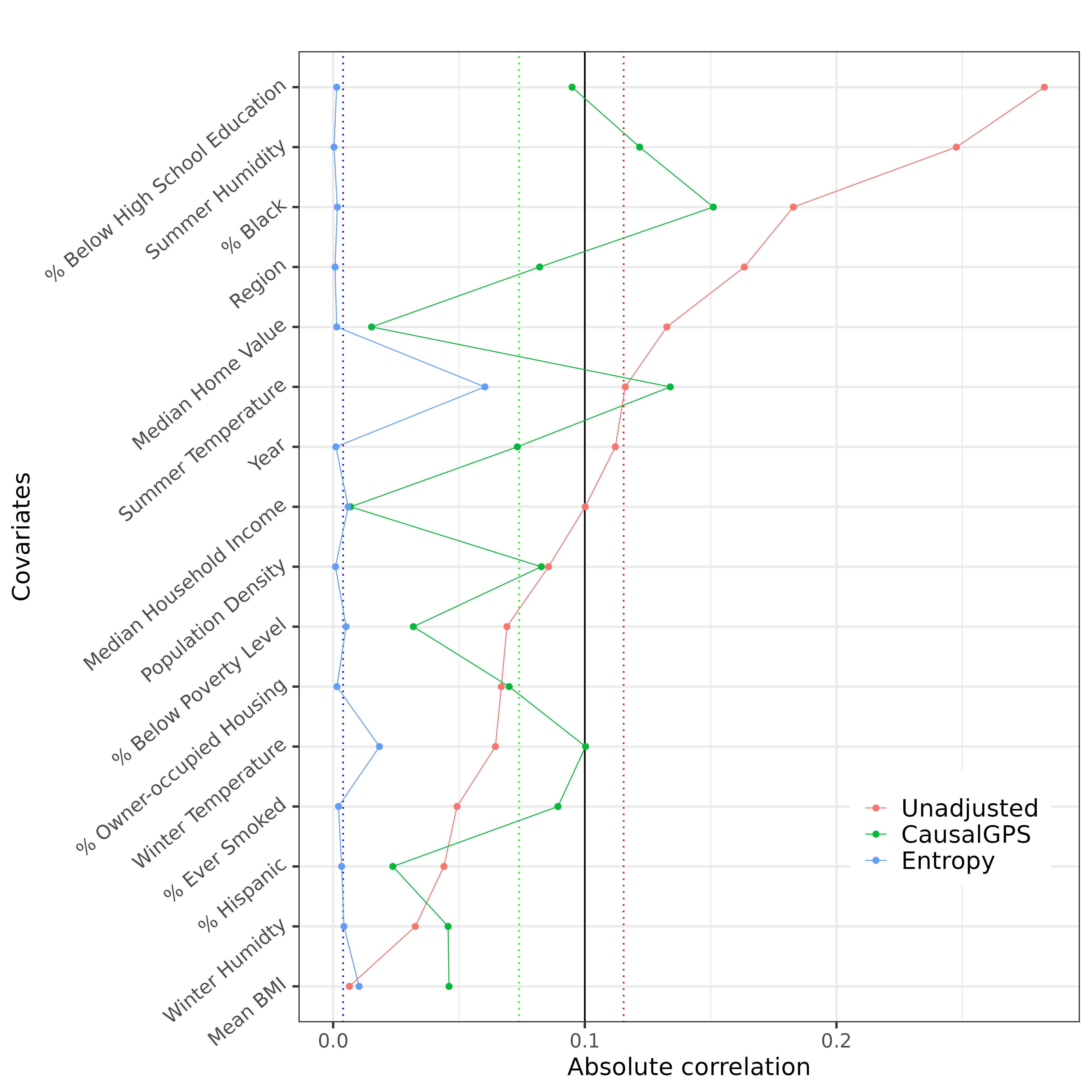}
    \caption{\textbf{Covariate balance plots of the absolute Pearson correlation coefficients in the unweighted Medicare data (unadjusted) and after weighing}. Absolute Pearson correlation between each covariates and $PM_{2.5}$ unadjusted (red), after CausalGPS matching adjustment (green), and after entropy weighting adjustment (blue). Absolute correlation of 0.1 (solid black line) or lower is considered empirically achieving covariate balance achieved \parencite{zhuBoostingAlgorithmEstimating2015}. Mean absolute correlation across all covariates is shown by the dotted line for unadjusted (red) CausalGPS (green) and entropy weighting (blue).}
\end{figure}

\begin{figure}[H]
    \centering
    \includegraphics[width=\textwidth]{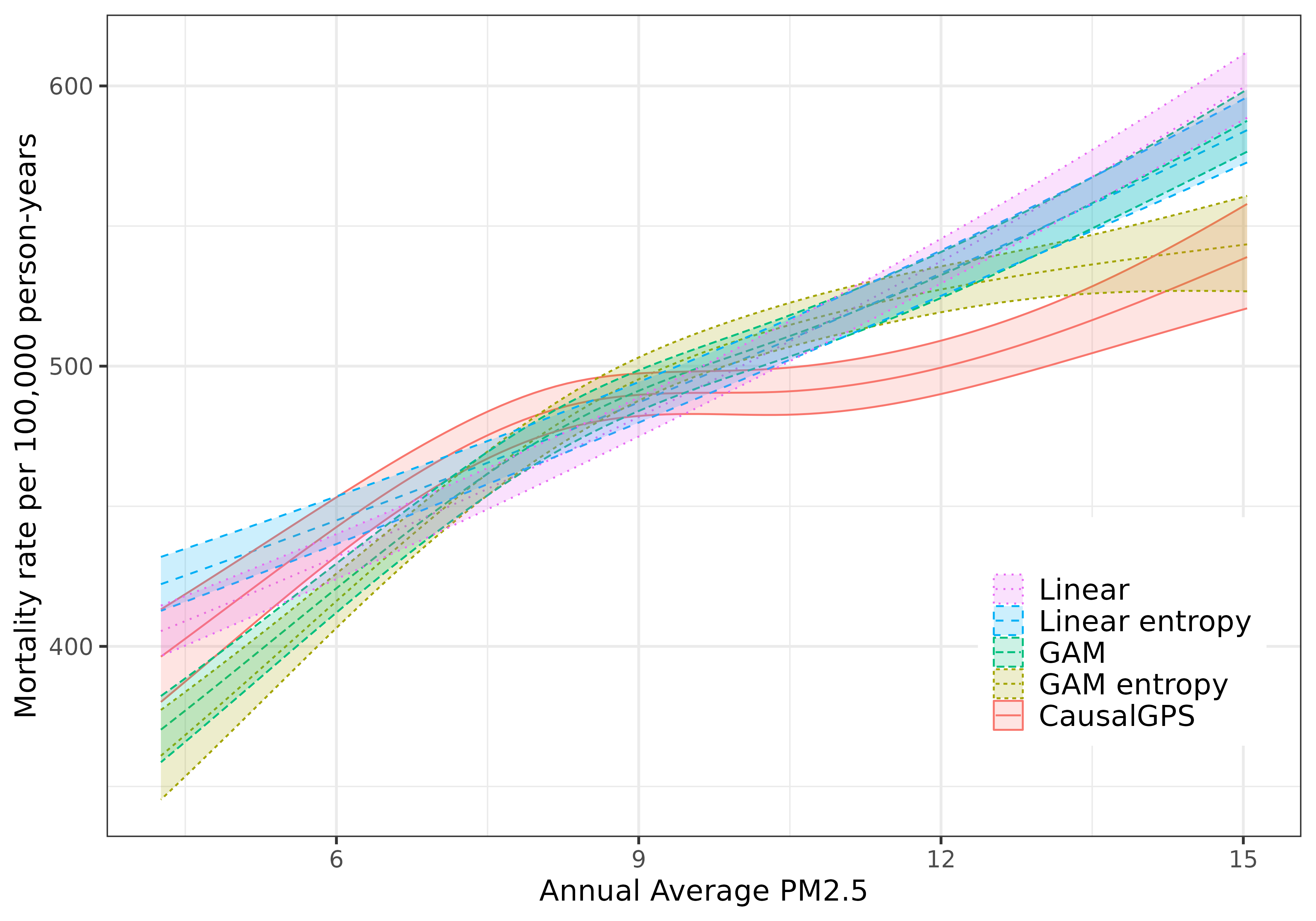}
    \caption{\textbf{Estimated ERC relating mortality as a function of annual average $PM_{2.5}$ on Medicare enrollee cohort with 95\% confidence intervals}. Estimator are designated by color; models failed to converge for change point and change point entropy models.}
\end{figure}

\begin{figure}[H]
    \centering
    \includegraphics[width=\textwidth]{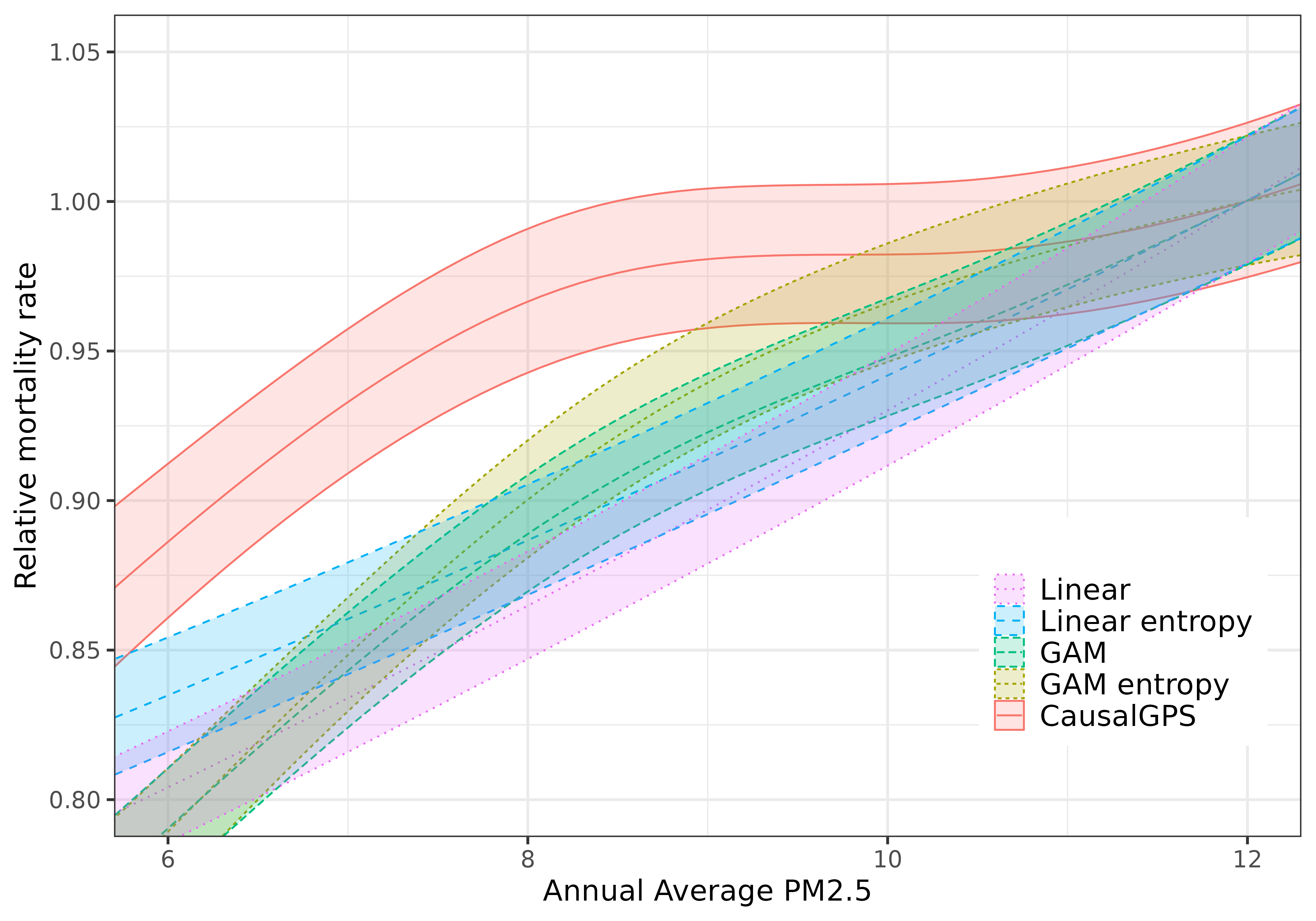}
    \caption{\textbf{Point estimates and 95\% confidence intervals of the relative mortality rate corresponding to decreases in annual average $PM_{2.5}$ with respect to 12 $\si{\micro\gram/\meter^3}$ on average for the Medicare population}. Estimator are designated by color; models failed to converge for change point and change point entropy models.}
\end{figure}

\newpage 

\section{Discussion}
Air pollution remains a significant threat to human health. Determining the shape of the exposure-response curve (ERC) between $PM_{2.5}$ exposure and all-cause mortality, while difficult to determine from large-scale observational studies, is crucial for informing policy decisions. Various approaches have been developed to confront this challenge, and our analysis represents one of the first comprehensive comparisons of these approaches under diverse data generating mechanisms that reflect real-world scenarios in air pollution epidemiology. Our analysis was designed to be accessible to the public, and the code for each step of the analysis is available on GitHub. Furthermore, we applied each method to data from the Medicare database, which represents the largest cohort to date encompassing the contiguous United States.

\subsection{Simulation}
We briefly summarize our findings and contextualize them within the existing literature. In our simulations, we initially examined commonly employed regression methods, including multivariate linear regression, generalized additive models with smooth terms for nonlinear effects, and change point models for detecting threshold effects. Each of these regression methods do not explicitly separate a design and analysis stage and are not formulated in a causal inference framework \parencite{imbensCausalInferenceStatistics2015}. Regression methods tend to perform well under two conditions: when the true ERC relationship is consistent with the regression model and there are no heterogeneous treatment effects (i.e., the model is correctly specified). For instance, when the true exposure-response relationship is linear ($\mu_{\text{linear}}$) the Linear estimator exhibits the lowest absolute bias and RMSE across all exposure scenarios. This finding is consistent with standard linear model theory, where the ordinary least squares estimator is considered the best linear unbiased estimator of the average treatment effect when the outcome model is correctly specified \parencite{lehmannTheoryPointEstimation2006}. Notably, even in the presence of nonlinear and interaction relationships between the confounders and the exposure ($E_{\text{nonlinear}}, E_{\text{interaction}}$), the linear model remains unbiased in our simulations. Similarly, the change point regression method exhibited low absolute bias and RMSE when the ERC takes the form of a threshold function ($\mu_{\text{threshold}}$). 

However, in our simulation, regression methods proved inadequate when heterogeneous treatment effects were present, particularly in cases involving nonlinear or interactive relationships between the exposure and confounders (e.g., $E_{\text{nonlinear}}, E_{\text{interaction}}$). When heterogeneous treatment effects were present, applying naive regression models often led to the largest absolute bias. Given the relatively low variance in regression methods, especially linear models, we often see them performing with comparable RMSE to causal inference methods at low sample sizes, even in the presence of substantial bias. While comparable RMSE might convey similar performance, an examination of the ERCs reveals that regression methods misrepresented the exposure-response relationship. This finding underscores the importance of caution when using single metrics to quantify ERCs. In the context of our simulation study, we demonstrate that regression methods are only appropriate if the ERC functional form is known and if there is no interaction between any of the covariates and the exposure in generating the outcome.

Our second class of estimators employed entropy-balancing weights in a linear model, generalized additive model, and change point model for the outcome. In our simulations, entropy weighting methods emerged as the optimal estimator in the presence of heterogeneous treatment effects when the exposure model takes the form of a linear function of the confounders ($E_{\text{linear}}$, $E_{\text{heavy tail}}$). Additionally, as the sample size in our simulation increased, the absolute bias and RMSE of entropy weighting estimators decreased. For example, in the linear ERC setting, we observed that weighting methods performed well in the presence of heterogeneity ($\mu_{\text{linear,int}}$) under $E_{\text{linear}}$ and $E_{\text{heavy tail}}$, exhibiting reduced bias compared to regression methods when the sample sizes exceeded 200. Larger sample sizes can allow for better covariate balance with a lower likelihood of generating extreme weights \parencite{tubbickeEntropyBalancingContinuous2022}. Extreme weights can introduce greater variability in the estimated ERC \parencite{leeWeightTrimmingPropensity2011}. In the setting of no heterogeneous effects, the weighting methods performed comparably to their regression counterparts. It is worth noting that entropy balancing weights tend to yield larger RMSE in these settings, likely due to the fact that balancing weights are constructed prior to regression, reducing bias but introducing variability into the estimates \parencite{golinelliBiasVarianceTradeoffs2012}.

Nonetheless, entropy balancing methods failed to produce unbiased results when the exposure followed a nonlinear function of confounders $E_{\text{nonlinear}}$, $E_{\text{interaction}}$. For example, in the linear ERC scenario, $\mu_{\text{linear}}$, the bias and RMSE increased for the entropy balancing methods compared to the unweighted regression methods. While entropy balancing effectively eliminates the correlation between the exposure and all covariates (Figure 2), this alone may not guarantee satisfactory covariate balance \parencite{yiuCovariateAssociationEliminating2018}. Removing the correlation between covariates and exposure eliminates the association between the covariate means and the continuous treatment, but it does not necessarily imply independence between the reweighted covariates and exposure. Balancing on higher moments of the covariates would likely enhance the performance of entropy balancing estimators and potentially correct the bias observed in the $E_{\text{nonlinear}}$ and $E_{\text{interaction}}$ settings \parencite{tubbickeEntropyBalancingContinuous2022}.

Finally, we evaluated the performance of the continuous matching causal inference method implemented in the CausalGPS package. Generally, the continuous matching estimator exhibited poor performance when the sample size was small. However, its performance improved significantly as the sample size increased, especially in the presence of heterogeneous treatment effects and complex exposure scenarios. Among the seven estimators, only CausalGPS performed well under the $E_{\text{nonlinear}}$ and $E_{\text{interaction}}$ scenarios with heterogeneous treatment effects. Nevertheless, CausalGPS demonstrated greater variability in ERC fit and exhibited higher RMSE values compared to other estimators. The superior performance of the CausalGPS method at larger sample sizes is not surprising, as it employs gradient boosting to specify the generalized propensity score (GPS). Gradient boosting allows for capturing nonlinearities and interactions when estimating the relationship between confounders and the exposure, and a larger sample size provides more data for training the machine learning algorithm, resulting in improved GPS approximation. Moreover, larger sample sizes enhance the CausalGPS algorithm's ability to identify appropriate matches for imputing potential outcomes.

We observed unexpected results in the sublinear ERC setting $\mu_{\text{sublinear}}$. Surprisingly, the GAM entropy model exhibited high bias and RMSE under heterogeneous treatment effects in the $E_{\text{linear}}$ and $E_{\text{heavy tail}}$ setting. We anticipated that the entropy balancing weights would reduce the bias of the regression estimators, but the bias persisted. One possible explanation for this observation is the placement of knots for the splines. In our study, the nonlinear component of the sublinear ERC occurs at lower levels of the exposure, while most GAM algorithms distribute knots at equally spaced quantiles of the data. Since we fixed the number of knots in our GAM estimator to four (refer to supplementary materials), it is likely that the knot placement is insufficient to capture the non-linearity that occurs at lower exposure levels. We may expect improved performance if we include more knots at the exposure values where we anticipate nonlinear behavior to occur.

Our findings underscore several important considerations for researchers interested in assessing the effect of a continuous exposure and generating an ERC. The first consideration pertains to the presence of heterogeneous treatment effects. We generally find that regression methods, when not incorporated within a causal inference framework, are unsuitable when anticipating heterogeneous treatment effects. In a regression setting, interactions are not accounted for in the model specification, whereas covariate balancing mitigates this dependency. However, if there is limited or no evidence of heterogeneous treatment effects, regression methods can be a viable framework, particularly at smaller sample sizes. As the sample size increases, we recommend adopting a causal inference framework, such as entropy balancing weights, to address potential confounding. Causal inference methods, such as weighting and continuous matching, separate the design and analysis stages, improving the objectivity of outcome data analysis \parencite{rubinObjectiveCausalInference2008}.

When using entropy balancing weights, it is important to carefully consider the balance of higher moments of the covariates and the interactions between covariates and the exposure to ensure the absence of residual confounding. In our analysis, we initially balanced only on the covariate means to eliminate correlation between the covariates and the exposure. However, we found this approach to be inadequate in the $E_{\text{nonlinear}}$ and $E_{\text{interaction}}$ scenarios. We propose a two-step process: first, balancing on the covariate means and then assessing remaining dependencies by estimating the ERC between the exposure and covariates using the generated weighting scheme. If the resulting ERC is completely flat and the derivative is zero, this indicates sufficient balance in the weighting scheme \parencite{tubbickeEntropyBalancingContinuous2022}. If this is not the case, we recommend incorporating additional covariate moments for balance. It is important to exercise caution when increasing the moments of balance as it can reduce bias but increase the variance of the estimator \parencite{tubbickeEntropyBalancingContinuous2022}. A suggestion from the literature is to balance two or three moments under a continuous treatment, which corresponds to balancing the covariate's mean, variance, and skew \parencite{vegetabileNonparametricEstimationPopulation2021}.

In the context of a large sample size where the functional form of the underlying ERC curve is unknown, we recommend employing the continuous matching framework provided by the CausalGPS package. The CausalGPS estimator demonstrates robustness against misspecification of either the GPS or outcome models, and larger sample sizes enable machine learning methods to more accurately estimate the GPS. Moreover, the matching step reduces the dependence between the exposure and potential confounders, resulting in causal effect estimates that are less influenced by choices made during outcome modeling \parencite{wuMatchingGeneralizedPropensity2022}. However, it is important to note that we observed considerable variability and inadequate covariate balance when utilizing the CausalGPS package with smaller sample sizes. As a result, we do not recommend using it in this setting.

\subsection{Data application}
In our data application, we conducted the first application of five statistical approaches to a data set containing over 500 million person-years of Medicare data, aiming to estimate the impact of annual $PM_{2.5}$ concentration on all-cause mortality (Figures 5, 6). Our findings reveal several noteworthy results. Each of the estimators demonstrated an increase in all-cause mortality in relation to annual average $PM_{2.5}$ concentration, consistent with prior research \parencite{joseyAirPollutionMortality2023, wuMatchingGeneralizedPropensity2022, wuEvaluatingImpactLongterm2020}. From our nonlinear estimators (GAM, GAM entropy, CausalGPS), we observed evidence of a nonlinear association between log mortality and $PM_{2.5}$, indicating the need to relax the linearity assumption. Moreover, based on the outcomes of our simulation study, the ERC generated by the GAM entropy and CausalGPS estimators seem most plausible, given the substantial sample size of our data, the presence of a nonlinear relationship between exposure and outcome, and the potential existence of heterogeneous treatment effects \parencite{joseyAirPollutionMortality2023}. Both the GAM entropy and CausalGPS estimated ERCs displayed a steep rise in the mortality rate at lower $PM_{2.5}$ concentrations, followed by a gradual plateau at higher concentrations. The GAM entropy and CausalGPS estimators exhibited a sublinear relationship between exposure and outcome at concentrations below 9 $\si{\micro\gram / \meter^3}$, though the ERC generated by the CausalGPS estimator leveled off at a lower concentration of annual average $PM_{2.5}$. Although the change point models failed to converge on this data set, the sublinear shape exhibited by the nonlinear causal inference methods at low concentrations suggests that a threshold model is unlikely. 

Regarding the relative mortality rate associated with a decrease in annual average $PM_{2.5}$ in relation to the current NAAQS annual standard of 12 $\si{\micro\gram / \meter^3}$, the CausalGPS package projected a more moderate reduction in relative mortality for concentrations ranging from 9-12 $\si{\micro\gram / \meter^3}$ compared to other estimators (Figure 6). Once again, considering our simulations, the GAM entropy and CausalGPS estimators appeared most plausible, providing varying magnitudes of evidence for establishing a lower NAAQS on annual average $PM_{2.5}$. The shape of of the relative mortality curve for CausalGPS is consistent with other results \parencite{joseyAirPollutionMortality2023} and offers evidence of a sublinear relationship between $PM_{2.5}$ and all-cause mortality. This analysis contributes further evidence in support of implementing stringent standards for $PM_{2.5}$ emissions, suggesting that a lower standard could prevent a significant number of premature deaths.

\subsection{Conclusion}
There are several limitations to our analysis. The primary limitation is that the evaluation of different estimators of the ERC was conducted using simulations. Our study was restricted to examining a limited number of scenarios, and it is possible that results may vary under different data-generating processes. However, we included various specifications of the ERC, covariates, and confounding scenarios that we believe represent a wide range of plausible situations. Another limitation is the choice of estimators. There are numerous estimators available in the regression and causal inference literature. We selected our estimators based on their availability in standard software and their widespread usage in practice.

Determining the causal relationship between a continuous exposure and outcome is a critical scientific endeavour. In the context of air pollution epidemiology, understanding the shape of the ERC between air pollution and adverse health effects has significant policy implications. Various modeling strategies have been employed to assess this relationship, and many researchers have advocated for the development and implementation of causal inference methods to inform air pollution policies \parencite{goldmanDonAbandonEvidence2019, petersPromotingCleanAir2019, caronePursuitEvidenceAir2020}. Although limited work has compared the performance of these different estimators, our analysis provides valuable guidance for researchers in choosing an appropriate method to estimate these consequential exposure-response curves accurately.

\section{Acknowledgements}
This research was supported by grants from the National Institutes of Health (Grant No. T32 ES007142, R01ES026217, R01MD012769, R01ES028033, 1R01ES030616, 1R01AG066793, 1R01ES029950, 1R01ES 034373-01) and the Alfred P. Sloan Foundation (Grant No. G-2020-13946, 1R01AG066793). We would like to thank these funding agencies for their support.

\newpage

\newrefcontext[sorting=nty]
\printbibliography[title={References}]

\end{document}